# Engineering Electrochromism in Ni-Deficient NiO through Defect, Dopant, and Strain Coupling

Katarina Jakovljević[1], Ana S. Dobrota[2], Igor A. Pašti[2,3*], Natalia V. Skorodumova[4]

[1] *5th Belgrade Gymnasium, Belgrade, Serbia*

[2] *University of Belgrade – Faculty of Physical Chemistry, Belgrade, Serbia*

[3] *Serbian Academy of Sciences and Arts, Kneza Mihaila 35, Belgrade, Serbia*

[4] *Applied Physics, Division of Materials Science, Department of Engineering Sciences and Mathematics, Luleå University of Technology, Luleå, Sweden*

**Abstract:**

The electrochromic response of Ni-deficient NiO is governed by vacancy-mediated electronic processes that can be strongly influenced by dopant chemistry and lattice deformation. Using density functional theory, we systematically investigated Cu-, Sn-, and V-doped Ni-deficient NiO(001) surfaces and examined alkali-ion insertion at surface Ni vacancies. Li insertion proceeds as nearly complete ionic electron donation (~+0.9 e), but the fate of the injected electron depends on dopant identity. V-doping preserves framework-dominated charge compensation and leads to conventional bleaching through filling of vacancy-associated hole states. In contrast, Sn actively traps the injected charge, generating dopant-assisted optical transitions and reversing the electrochromic response, while Cu produces an intermediate spectral redistribution without significant dopant reduction. Substitution of Li by Na or K in the V-doped system does not alter the switching mechanism, confirming that vacancy-state filling governs the optical behavior. Biaxial tensile strain enhances the energetics of Li insertion but reduces optical contrast by altering the defect electronic structure. These results establish dopant activity, vacancy stabilization, and lattice strain as key parameters controlling electrochromism in NiO-based materials.

## 1. Introduction

Introduction

Electrochromic (EC) materials reversibly and persistently change optical properties (transmittance/absorbance/reflectance) under an applied electrical bias, enabling actively tunable glazing and coatings rather than static heat/light management.[1] EC "smart windows" built from complementary EC layers, classically a cathodically coloring tungsten oxide ($WO_3$-based) film and an anodically coloring nickel oxide ($NiO_x$-based) film separated by an ion-conducting electrolyte, can modulate visible transmission while maintaining good optical quality and long-term reversibility.[1] Because windows strongly influence daylighting, glare, and solar heat gain, such adaptive control is widely viewed as an important route towards lowering operational energy demand and improving indoor comfort in buildings.[1] Within this device family, anodically coloring $NiO_x$ remains a critical "balancing" layer: it provides complementary optical absorption to $WO_3$, contributes to charge balance, and can be engineered as both an electrochromic and ion-storage component.[1] At the same time, $NiO_x$ is often considered the weaker link in terms of achievable optical contrast and, especially, long-term stability across different electrolytes and potential windows. These limitations drive continued materials design efforts (microstructure control, defect engineering, and doping).[2]

---

[*] *E-mail: igor@ffh.bg.ac.rs*

NiO is a wide-band-gap, typically p-type oxide whose non-stoichiometry and defect chemistry are central to its electrochemical and optical response.[3] In aqueous alkaline electrolytes, the dominant mechanistic picture links anodic coloring to proton-coupled redox processes at/near the surface, often rationalised via transformations between hydroxide and oxyhydroxide-like surface phases (the Bode-type scheme), with optical absorption associated with oxidized nickel states.[4] X-ray photoelectron spectroscopy-based analyses of hydrated $NiO_y$ (and related Ni–V oxides) further support increased populations of higher-valence nickel species in colored films and motivate extensions of simplified $Ni(OH)_2$/NiOOH-only schemes to include participation of $NiO/Ni_2O_3$-like components.[5] However, NiO electrochromism in aprotic, $Li^+$-conducting electrolytes can differ qualitatively. In particular, surface-controlled charge exchange involving both cations and anions has been argued to dominate in $LiClO_4$/propylene carbonate (PC), and the degree of modulation can depend strongly on the exposed surface facets and on the oxygen content x in $NiO_x$, reinforcing the notion that the active electrochromic volume is often confined to surface-proximal regions of porous grains.[6] Consistent with this electrolyte sensitivity, porous sputtered NiO films can deliver large optical modulation in both KOH and $LiClO_4$/PC, yet exhibit markedly different stress evolution and durability: for example, significant degradation can appear after ~1000 cycles in Li-PC under wide potential excursions, even when KOH cycling remains stable over far more cycles under the chosen conditions.[7] More generally, long-term cycling is often accompanied by progressive charge density loss (and hence reduced optical modulation), which has been modelled empirically using power-law and related decay descriptions.[7]

Alongside these established experimental frameworks, a complementary atomistic viewpoint has emerged for Ni-deficient NiO: the formation of Ni vacancies can generate hole polarons localized primarily on oxygen 2p states adjacent to the vacancy, and electrochromic switching can arise from filling/emptying these vacancy-derived hole states rather than invoking a simple $Ni^{2+}/Ni^{3+}$ transition.[8] In our previous DFT+$U$ study, the same qualitative picture was obtained in bulk and on the NiO(001) surface, and Li/Na/K incorporation into a Ni surface vacancy acted effectively as electron injection that quenches vacancy-related absorption by annihilating hole states.[8] This vacancy-centered, oxygen-polaron framework provides a useful starting point for disentangling how specific defects act as "color centers" and how charge-compensation pathways control optical modulation.[8]

Doping is one of the most widely used modification strategies to improve NiO electrochromic performance, complementing morphology control (nanostructuring, porosity), compositing, and electrolyte engineering.[2] Experimentally, reported benefits of doping span multiple metrics (optical modulation, switching speed, charge capacity, and cycling stability) and are commonly attributed to combinations of (i) altered microstructure/surface area, (ii) modified defect populations (cation/oxygen vacancies, interstitials), and (iii) improved electronic transport.[2] Representative examples illustrate both the promise and the interpretive ambiguity. Co-doped NiO nanoflake array films showed very high bleached-state transmittance and enhanced electrochromism compared with undoped analogues (including large Δ$T$ at 550 nm and improved switching), with proposed contributions from reduced crystallization, favorable nanoflake geometry, and increased p-type conductivity.[9] Vanadium doping has likewise been reported to refine grains, increase specific surface area, and accelerate $OH^-$ diffusion in porous NiO, enabling large transmittance modulation, fast response, and high cycle retention in alkaline operation.[10] Beyond transition-metal substitution, defect-engineering dopants such as Mg can "loosen" the film microstructure and increase oxygen-related defect concentrations, thereby improving charge transfer and increasing transmittance modulation in $NiO_x$.[11] More recent work has expanded doping beyond simple cation substitution. For instance, amorphous, porous C/N-doped NiO has been proposed for EC smart-window applications, demonstrating substantial optical modulation, consistent with the broader expectation that amorphization and porosity can facilitate ion access and accelerate switching.[12] At the device level, Zn-doped NiO has been used in multifunctional concepts that couple electrochromism with energy storage (panchromatic EC behavior and "EC-supercapacitor" integration), demonstrating how dopant selection can be leveraged in systems-oriented designs.[13] Finally, manufacturing-relevant studies increasingly combine dopants with binders/additives to enable scalable processing while retaining cycling stability, e.g., Zn-doped NiO combined with ethyl cellulose and In/Sn metal powder additives for large-area fabrication.[14] These results collectively demonstrate that "doping improves NiO" is empirically true in many cases, yet the mechanistic origins are often discussed in broad terms (porosity, conductivity, more active sites), leaving unresolved which electronic states are introduced/removed by a dopant and whether injected charge is compensated by Ni-derived redox, dopant redox, or vacancy-derived hole states.[2]

In realistic doped, Ni-deficient $NiO_x$ films, multiple defect motifs likely coexist: dopant–vacancy complexes, isolated Ni vacancies away from dopants, and dopant sites without nearby vacancies. This implies that the measured optical response can be expressed as a superposition of contributions from chemically distinct "color centers", each with its own charge-compensation pathway and spectral fingerprint.[2] An atomistic, bond-level description is therefore needed to connect: (i) *where* the compensating electron goes during cation insertion (vacancy-centered oxygen holes vs dopant-centered states), and (ii) *how* that localization modifies the electronic transitions responsible for visible absorption.[8] Motivated by the vacancy-polaron framework established in ref.[8], the present work extends that approach to doped, Ni-deficient NiO(001) in order to test, under controlled, comparable conditions, how different dopant chemistries compete with vacancy-derived states during alkali insertion and how this competition governs electrochromic contrast.[8] We focus on dopants that (i) are widely used experimentally to improve NiO electrochromism, and (ii) span distinct electronic behaviors (spectator-like *vs* redox-active participation), thereby enabling a systematic evaluation of whether improved macroscopic performance correlates with dopant participation in charge compensation or predominantly with dopant-induced reorganization of vacancy states.[2] In addition, because practical NiO films are often subject to epitaxial mismatch and residual stress (especially on transparent conductors), we also consider strain as a device-relevant parameter that can perturb defect energetics and optical transitions even when the underlying charge-transfer mechanism remains intact.[7]

## 2. Computational details

First-principles density functional theory (DFT) calculations were performed using the Vienna *Ab initio* Simulation Package (VASP).[15–18] The exchange-correlation energy was described within the Generalized Gradient Approximation (GGA) using the Perdew-Burke-Ernzerhof (PBE) functional,[19] in combination with the projector augmented wave (PAW) method.[20] A plane-wave kinetic energy cutoff of 800 eV was employed throughout. Electronic occupations were treated using Gaussian smearing with a width of $\sigma = 0.025$ eV. On-site Coulomb interactions were included within the DFT+$U$ framework using the Liechtenstein approach.[21] The Hubbard correction was applied to both the Ni 3d ($U = 6.5$ eV, $J = 1$ eV) and O 2p states ($U = 9$ eV, $J = 1$ eV). Non-spherical contributions from the gradient corrections inside the PAW spheres and to the Kohn-Sham potential were explicitly considered. Spin-polarized calculations were performed in all cases.

The NiO(001) surface was modeled using a four-layer slab constructed in a p(2×2) supercell containing 64 atoms. The two bottom layers were fixed at bulk positions during structural relaxation, while the two top layers were allowed to fully relax. A Monkhorst-Pack k-point mesh was used for geometry optimizations,[22] with a 5×5×1 grid employed for surface calculations. Density of states (DOS) calculations were performed using the Blöchl tetrahedron method.[23] The Ni-deficient surface was generated by removing one Ni atom from the outermost surface layer. Doped Ni-deficient surfaces were constructed by substitutionally replacing a surface Ni atom with Cu, Sn, or V within the same p(2×2) slab, maintaining identical supercell size and defect concentration. All structural optimizations were performed until residual forces on each atom were below 0.01 eV Å$^{-1}$, while electronic self-consistency was achieved to an energy tolerance of 10$^{-5}$ eV. The NiO(001) surface was selected because it is densely packed, non-polar, and does not undergo surface reconstruction, thereby providing a structurally well-defined platform for systematically analyzing vacancy-induced hole localization and its modulation by dopant incorporation. The binding energies of alkali metal atoms ($E_b$) were calculated with respect to the energy of an isolated atom ($E_A$), as $E_b = E_{surf+A} - E_{surf} - E_A$, where $E_{surf+A}$ and $E_{surf}$ stand for the energy of the NiO surface with bonded metal atom A, and the energy of the considered NiO surface. Optical spectra were calculated using the same methodology as in ref. [8]

## 3. Results and Discussion

### 3.1. Doping NiO(001) and $V_{Ni}$ formation in doped NiO(001)

The calculated energetics reveal clear dopant-dependent site preferences that are further modified by the presence of a surface Ni vacancy. On the stoichiometric NiO(001) surface, Cu and Sn preferentially substitute Ni atoms in the outermost layer, being stabilized at the surface by 0.63 eV (Cu) and as much as 3.28 eV (Sn) relative to subsurface substitution. In contrast, V exhibits a slight preference for the subsurface position, which is 0.10 eV lower in energy than the surface site. This trend reflects the balance between coordination and relaxation effects: Cu and, particularly, Sn benefit from reduced coordination and enhanced structural flexibility at the surface, whereas V is better stabilized in a more bulk-like environment with higher coordination.

The introduction of a surface Ni vacancy substantially alters this energetic landscape. For all three dopants, the globally most stable configuration corresponds to the dopant occupying a surface site with the vacancy positioned at the next-nearest-neighbor (NNN) distance. Relative to this minimum-energy configuration, alternative vacancy placements are energetically penalized by 0.42-0.96 eV for Cu, 0.19-0.90 eV for Sn, and 0.56-1.40 eV for V. In particular, the nearest-neighbor (NN) arrangement is consistently less favorable, indicating that direct dopant-vacancy adjacency does not maximize stabilization. Instead, excessive local lattice distortion and electronic overcompensation likely destabilize the NN geometry. At larger separations, the dopant-vacancy interaction weakens, and the stabilization diminishes. The NNN configuration thus represents an optimal compromise between elastic relaxation and electronic coupling. This systematic preference for the NNN geometry suggests that dopant stabilization is mediated not merely by structural strain relief, but also by modulation of the vacancy-associated electronic states. Since Ni vacancies in NiO induce localized hole states predominantly on neighboring oxygen atoms, the energetic minimum at intermediate dopant–vacancy separation indicates that the dopant perturbs and stabilizes these O-derived hole states most efficiently without directly competing for charge localization.

Bader charge analysis (Henkelman partitioning) provides a consistent picture of substantial charge transfer from all dopants to the oxide lattice for the most stable NNN configurations with surface substitution. Using the isolated-atom valence electron counts (Cu: 12, Sn: 4, V: 5) as references, the Bader electron populations of 9.9645 (Cu), 1.9517 (Sn), and 3.1279 (V) correspond to net electron donation of 2.04 e, 2.05 e, and 1.87 e, respectively. Although Bader charges cannot be mapped one-to-one onto formal oxidation states, these values clearly indicate that the dopants adopt a strongly cationic character close to an effective +2 charge when substituting surface Ni sites in Ni-deficient NiO(001). This is fully consistent with the NiO host environment dominated by $Ni^{2+}$–$O^{2-}$ ionicity and implies that dopant incorporation does not introduce an extreme donor/acceptor imbalance (e.g., Sn behaving as a +4 donor). Instead, Cu, Sn, and V primarily act as ~divalent substitutional cations, so their main role is to tune the local electrostatics and bonding around the Ni vacancy

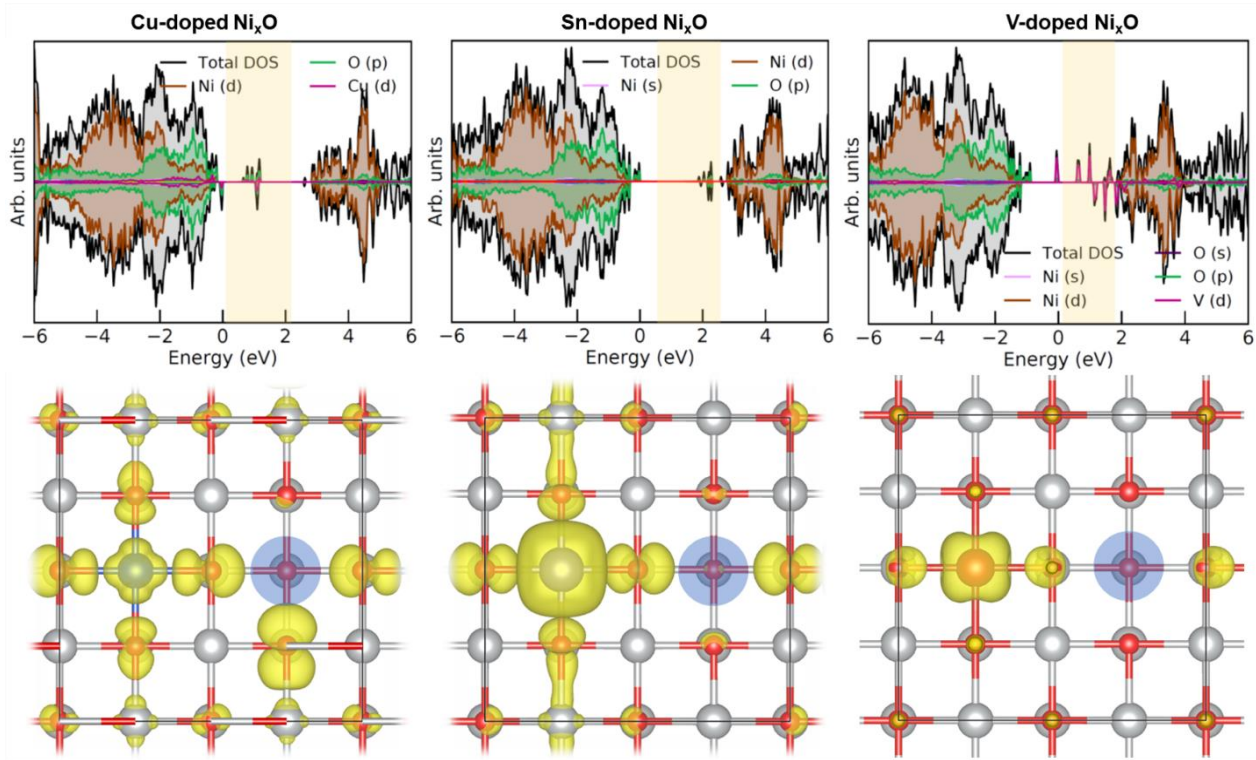

**Figure 1.** Projected density of states (PDOS) for Cu-, Sn-, and V-doped Ni-deficient NiO(001) in the most stable NNN configuration with the dopant located in the surface layer (top panels). The shaded region indicates the energy window used to construct the partial charge densities. The lower panels show the corresponding partial charge densities obtained by integrating the unoccupied states within the selected energy range. Yellow isosurfaces represent the spatial distribution of these states. The blue circle marks the position of the surface Ni vacancy.

The projected density of states (**Figure 1**, top row) and the partial integrated local density of states (**Figure 1**, bottom row) reveal that dopant incorporation qualitatively modifies the spatial character of the vacancy-induced unoccupied states. In the Cu-doped system, although the DOS in the selected energy window remains largely dominated by O(p) contributions, the real-space localization shows a clear participation of the Cu atom, indicating that the vacancy-derived hole state is partially redistributed onto the dopant through Cu–O hybridization. This redistribution becomes significantly more pronounced in the Sn-doped surface, where the empty-state density exhibits strong localization centered on the Sn atom and its coordinating oxygen atoms. The vacancy-associated defect state is therefore no longer confined to the oxygen sublattice but acquires substantial dopant character. The effect is most pronounced for V doping: both the DOS and the spatial charge density show strong V(d) contributions within the defect manifold, with the unoccupied states clearly localized on the V site itself. Taken together, these results establish a systematic trend in dopant participation in the vacancy-induced electronic structure. While the Ni vacancy remains the origin of the defect states in all cases, the degree of dopant involvement increases from Cu to Sn to V, progressively transforming the hole state from predominantly oxygen-centered to increasingly dopant-hybridized. This demonstrates that substitutional doping does not merely perturb the lattice energetics but actively reshapes the electronic structure of the vacancy manifold at the surface.

### 3.2. Electrochromism of doped NiO(001)

The calculated Li binding energies (**Table 1**) clearly demonstrate that lithiation proceeds preferentially via occupation of the Ni vacancy rather than adsorption at the dopant site for all doped Ni-deficient NiO(001) surfaces investigated. In the Cu-doped system, Li binding at the vacancy site is strongly exothermic (−5.32 eV), exceeding the stability of Li at the Cu site (−2.75 eV) by 2.57 eV. The preference is even more pronounced for Sn and V: Li binds weakly on the dopant sites (−0.49 eV for Sn and −0.32 eV for V), while insertion into the Ni vacancy remains strongly favorable (−4.45 and −3.61 eV, respectively), making the vacancy site more stable by 3.96 eV (Sn) and 3.29 eV (V). These results establish the surface Ni vacancy as the dominant thermodynamic host site for Li across all dopants, consistent with lithiation acting as electron injection into the vacancy-associated defect manifold. At the same time, the magnitude of Li stabilization at the vacancy decreases in the order Cu > Sn > V, indicating that dopant identity modulates the local electronic structure of the vacancy environment and its ability to stabilize the inserted Li$^+$.

**Table 1.** Energetics of lithiation on doped Ni-deficient NiO(001): Li binding energies at surface dopant sites and Ni vacancy ($V_{Ni}$) positions in the most stable NNN configuration.

| Surface | $E_b$(Li) / eV | |
| --- | --- | --- |
| | Dopant site | $V_{Ni}$ site |
| Cu-doped | −2.75 | −5.32 |
| Sn-doped | −0.49 | −4.45 |
| V-doped | −0.32 | −3.61 |

Bader charge analysis using the Henkelman partitioning scheme confirms that lithiation proceeds predominantly as ionic Li insertion into the surface Ni vacancy in all doped systems. Taking the isolated Li valence electron count as reference, the Bader populations of Li in the vacancy (2.102–2.104 e) correspond to an electron donation of approximately 0.90 e, indicating that Li adopts an effective charge close to +0.9 irrespective of dopant identity. Thus, the lithiation step itself is uniformly cationic across Cu-, Sn-, and V-doped surfaces. However, the spatial redistribution of the donated electron is strongly dopant-dependent (**Figure 2**). In the Cu-doped surface, the Bader population of Cu changes only marginally upon lithiation (9.9645 → 9.9268; Δ = −0.038 e), demonstrating that Cu does not act as the primary electron acceptor. Instead, the injected electron is redistributed over the oxide framework, as confirmed by the three-dimensional charge redistribution maps, which show that the dominant charge accumulation remains localized on oxygen sites surrounding the vacancy. Cu therefore behaves largely as a structural modifier that perturbs the defect energetics but does not significantly participate in the redox process.

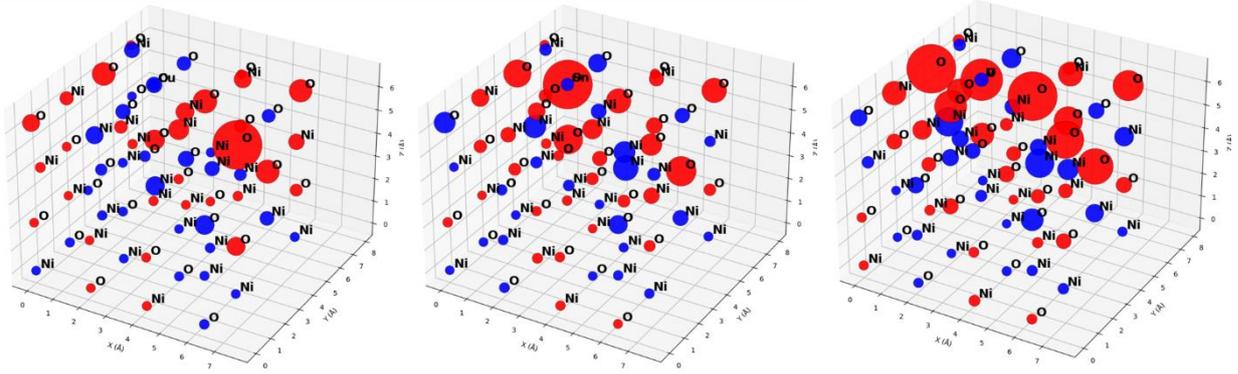

**Figure 2.** Three-dimensional visualization of Bader electron-count changes (Δe) upon lithiation of doped Ni-deficient NiO(001) surfaces. The atomic positions correspond to the relaxed slab structure. Spheres are centered at atomic sites and scaled proportionally to the absolute magnitude of Δe. Red spheres indicate positive Δe (electron accumulation, atom becomes more negative), while blue spheres indicate negative Δe (electron depletion, atom becomes more positive). Note that the scaling factors for sphere sizes differ across the figures presented. Ni, O, and dopant atoms (Cu/Sn/V) are labeled explicitly. The dopant is located in the surface layer, and Li is inserted into the surface Ni vacancy. The spatial distribution highlights the dopant-dependent redistribution of the injected electron: charge accumulation remains largely oxygen-centered in the Cu-doped surface, while significant dopant-centered accumulation is observed for Sn and, to a lesser extent, V.

In contrast, Sn exhibits a substantial increase in Bader population upon lithiation (1.9517 → 2.3714; Δ = +0.420 e), indicating pronounced partial reduction of the dopant. The 3D Δe visualization clearly shows enhanced electron accumulation on the Sn site and its immediate oxygen coordination shell, demonstrating that a considerable fraction of the injected charge is directly accommodated by Sn-derived states. This dopant-centered charge trapping modifies the electronic structure near the vacancy and prevents a complete return to a purely oxide-dominated electronic picture. The V-doped surface displays intermediate behavior. The Bader population of V increases from 3.1279 to 3.2730 (Δ = +0.145 e), confirming partial dopant involvement in the reduction process. The corresponding Δe distribution reveals mixed charge accommodation: a portion of the injected electron localizes on V, while a significant fraction remains distributed over neighboring oxygen atoms. Thus, V participates in the electronic compensation, but less strongly than Sn.

Comparison of the density of states and real-space charge distributions before (**Figure 1**) and after (**Figure 3**) lithiation shows that insertion of Li largely neutralizes the vacancy-induced hole states in all systems. In the Cu-doped case, the vacancy-related unoccupied states near the Fermi level almost completely disappear after lithiation, consistent with electron filling of O-centered defect states and minimal Cu participation. For Sn, although vacancy states are reduced, dopant-centered contributions remain visible, reflecting direct involvement of Sn orbitals in accommodating the injected charge. The V-doped surface again exhibits intermediate behavior, with partial filling of vacancy states and residual dopant-related features.

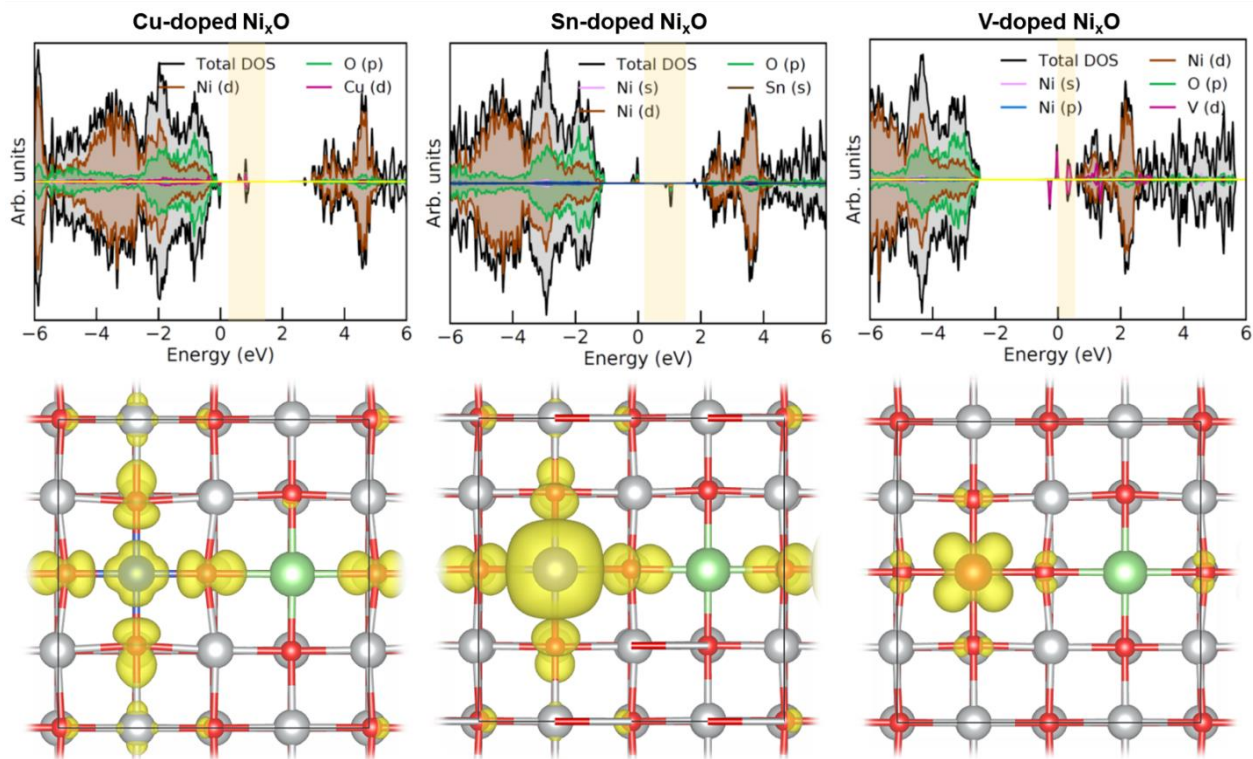

**Figure 3.** Projected density of states (PDOS) and spatial distribution of the unoccupied states for Cu-, Sn-, and V-doped Ni-deficient NiO(001) surfaces before and after lithiation (Li inserted into the Ni vacancy). The shaded region in the DOS panels indicates the energy range used to construct the corresponding charge density plots shown below. Yellow isosurfaces represent the spatial localization of these states. The green sphere marks the position of the inserted Li atom at the former Ni vacancy site.

The optical spectra (**Figure 4**) demonstrate that dopant identity fundamentally modifies the lithiation-induced response of Ni-deficient NiO. In undoped Ni-deficient NiO, lithiation leads to bleaching because the injected electron fills vacancy-associated hole states, suppressing low-energy optical transitions. The V-doped surface follows this behavior closely: upon Li insertion into the vacancy, the absorption coefficient decreases across the visible range. This is consistent with the Bader analysis ($\Delta q(V) \approx +0.145$ e), which indicates only a modest dopant reduction, and with the spatial charge redistribution, which shows that the injected electron remains largely accommodated by the oxide framework. Thus, the electronic structure after lithiation approaches the vacancy-neutralized, oxide-like limit. In contrast, the Sn-doped system exhibits an inverted response, with lithiation increasing the absorption coefficient. This behavior correlates directly with the pronounced involvement of Sn in the electronic compensation process. The substantial increase in Sn Bader population ($\Delta q(Sn) \approx +0.420$ e) and the localized charge accumulation around the Sn–O coordination shell indicate that a significant fraction of the injected electron is trapped in dopant-derived states. Rather than simply eliminating vacancy-hole transitions, lithiation stabilizes new or enhanced dopant-assisted electronic transitions, thereby increasing optical absorption. The Cu-doped surface displays intermediate behavior, with lithiation reducing absorption in part of the spectrum while enhancing it in another region. Bader analysis shows a negligible reduction in Cu ($\Delta q(Cu) \approx -0.038$ e), and the spatial $\Delta e$ maps (**Figure 2**) confirm that the injected electron is primarily redistributed among the oxygen atoms surrounding the vacancy. Thus, Cu does not act as an electron sink. The non-monotonic spectral response instead reflects a redistribution of transition intensity within the oxide defect manifold: vacancy-state filling suppresses some low-energy transitions, while Cu-induced modification of the local electronic structure alters hybridization and oscillator strengths in other spectral regions.

Thus, although Li insertion consistently occurs at the Ni vacancy in all systems and proceeds as a nearly ionic process (Li ≈ +0.9), the optical outcome is governed by how the injected electron is accommodated. When compensation remains framework-centered (V and largely Cu), lithiation promotes bleaching. When the dopant actively participates in charge trapping (Sn), lithiation generates dopant-assisted transitions and enhanced coloration. The competition between framework-dominated and dopant-centered charge accommodation, therefore, determines whether lithiation results in bleaching, coloration, or a non-monotonic redistribution of optical intensity.

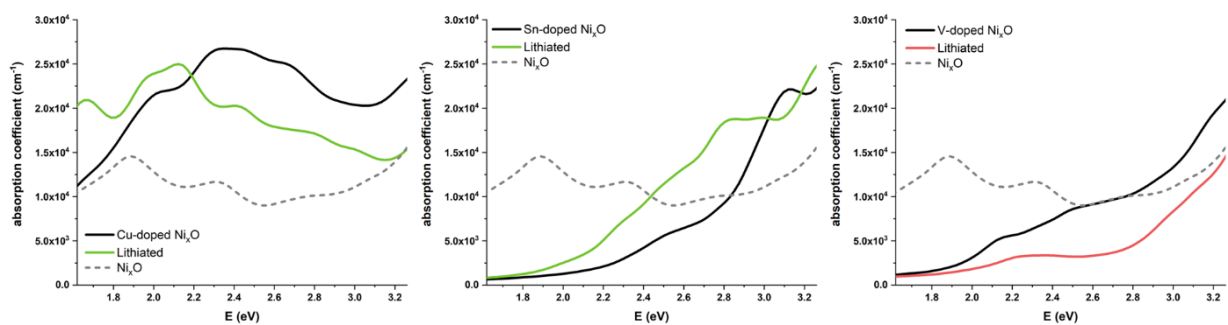

**Figure 4.** Calculated optical absorption coefficients of Cu-, Sn-, and V-doped Ni-deficient NiO(001) surfaces in the non-lithiated and lithiated states (Li inserted into the Ni vacancy). The dashed curve represents the absorption spectrum of undoped Ni-deficient NiO for comparison. Lithiation induces dopant-dependent changes in the visible absorption spectrum, leading to bleaching, enhanced coloration, or spectral redistribution, depending on the dopant identity. Note that pristine NiO(001) is practically transparent in the energy window shown.[8]

### 3.3. Cation effects on V-doped NiO(001)

Given the markedly dopant-dependent optical response identified for Li insertion, we selected the V-doped Ni-deficient NiO(001) surface for further investigation with larger alkali ions. This choice is motivated by the fact that V-doping preserves the "normal" electrochromic behavior of Ni-deficient NiO, i.e., lithiation leads to bleaching (**Figure 4**), in contrast to the inverted response found for Sn and the non-monotonic spectral redistribution observed for Cu. Consistent with this, our Bader analysis for Li indicated only modest dopant participation (ΔV ≈ +0.145 e), implying that charge compensation remains largely framework-dominated and therefore closely follows the vacancy-state filling mechanism established for undoped Ni-deficient NiO. Moreover, such electrochromic behavior has been experimentally observed, with switching to the bleached state occurring at low electrode potentials, i.e., under reductive conditions.[10] Having identified V-doping as the most representative case, we next replaced Li by Na and K at the surface Ni vacancy in order to assess how alkali-ion size and binding strength affect insertion thermodynamics and the ionic character of the inserted species. Bader charge analysis confirms that Na and K insertion into the Ni vacancy remains predominantly ionic. Using the isolated valence electron counts as reference, the Bader electron populations of Na and K in the vacancy correspond to net electron donations of 0.857 e (Na) and 0.847 e (K), i.e., effective charges of approximately +0.86 and +0.85. Compared to Li in the same vacancy site (≈+0.90), both Na and K are slightly less ionized, consistent with the weaker stabilization of the donated electron in the local defect environment for larger alkali ions. This trend is mirrored in the insertion energetics: Na binds strongly to the vacancy ($E_b$ = −2.98 eV), whereas K is substantially less stabilized ($E_b$ = −1.34 eV). Thus, while all three alkali ions insert as cations and donate nearly one electron to the defective oxide, increasing ionic size decreases both the degree of charge transfer and, more prominently, the thermodynamic driving force for vacancy occupation, with K showing the weakest binding among the tested alkali species.

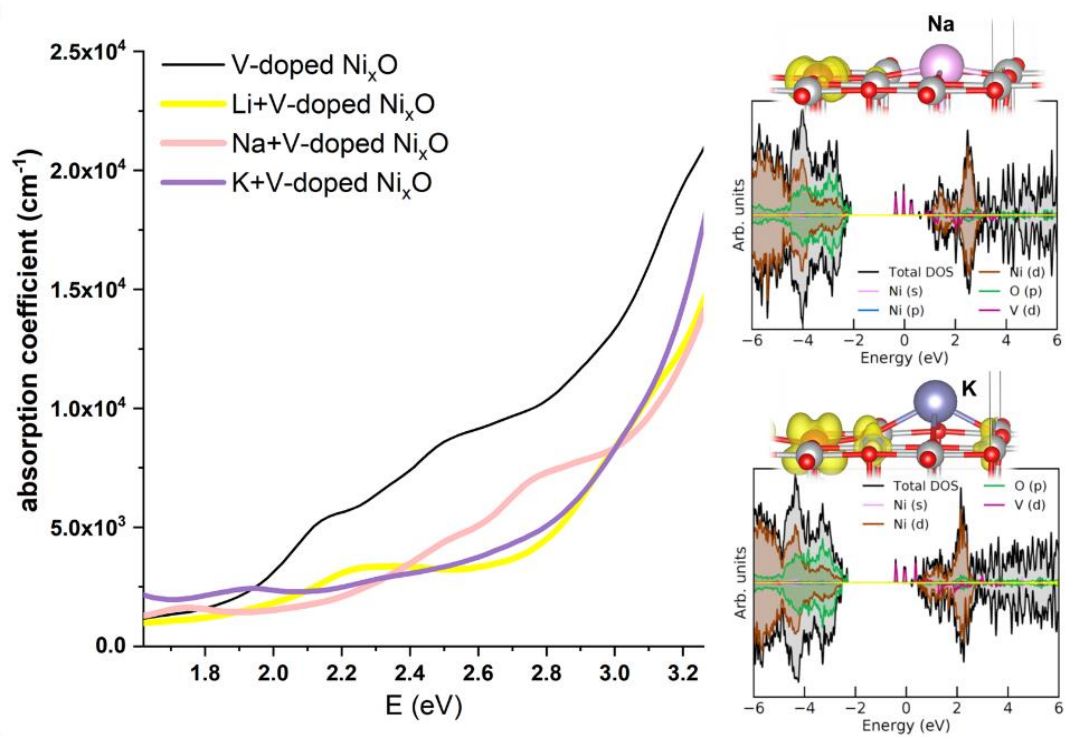

**Figure 5**. Calculated optical absorption spectra of V-doped Ni-deficient NiO(001) before and after alkali-ion insertion at the surface Ni vacancy. The black curve corresponds to the pristine V-doped surface, while the yellow, pink, and violet curves represent Li-, Na-, and K-inserted systems, respectively. Insets show the total and projected densities of states and the spatial distribution of the integrated unoccupied states above the Fermi level for representative Na- and K-inserted configurations.

The optical response of the V-doped Ni-deficient NiO(001) surface is remarkably insensitive to the identity of the inserted alkali ion. The insertion of Li, Na, or K into the surface Ni vacancy produces essentially the same spectral evolution (**Figure 5**), characterized by a pronounced reduction of the absorption coefficient in the visible region. This uniform bleaching behavior demonstrates that the electrochromic response is primarily governed by electron injection into vacancy-associated hole states rather than by alkali-specific chemical interactions. Bader charge analysis confirms that all three ions behave as strong electron donors, donating nearly 1 electron to the defective oxide framework. The integrated density of unoccupied states above the Fermi level shows that these vacancy-derived hole states are effectively neutralized upon insertion of any of the three alkali species, directly correlating with the observed suppression of low-energy optical transitions. Although Na and especially K relax further outward from the surface due to their larger ionic radii and exhibit weaker binding energies compared to Li (**Figure 5**), these geometric differences do not alter the fundamental electronic compensation mechanism. In the V-doped system, where dopant participation in the reduction process is modest, the electrochromic switching is controlled by vacancy-state filling within the oxide framework. Consequently, the bleaching response is robust with respect to alkali-ion size, confirming that electron donation to the defect manifold, rather than specific cation chemistry, dictates the optical behavior in this case.

### 3.4. Strain effects on electrochromism

Having established that V-doped Ni-deficient NiO(001) exhibits a robust "normal" electrochromic response, where alkali-ion insertion into the surface Ni vacancy consistently neutralizes vacancy-associated hole states and leads to bleaching, we next examine how mechanical deformation modulates this behavior. Strain is a particularly relevant control parameter for thin-film electrochromic oxides, where epitaxial mismatch with substrates, thermal expansion differences, and electrochemical cycling can impose appreciable biaxial stress. Because the switching mechanism in V-doped Ni-deficient NiO is governed by a balance between vacancy stabilization, electron donation from the inserted cation, and the degree of dopant participation in electronic compensation, biaxial strain is expected to perturb the local bonding environment and electronic structure of the vacancy complex. We therefore applied biaxial strain in the (001) plane and quantified its impact on the thermodynamics of Li insertion, the ionic character of the inserted species, and the change in the V charge state, in order to assess whether strain can be used as an additional handle to tune electrochromic energetics while preserving the desired bleaching mechanism.

Application of biaxial tensile strain in the (001) plane modulates the thermodynamics of Li insertion into the surface Ni vacancy of V-doped Ni-deficient NiO, while preserving the fundamental charge-compensation mechanism (**Table 2**). Moderate tensile strain enhances Li stabilization, with the binding energy becoming more negative at 1.25% strain (−4.02 eV) compared to the unstrained surface (−3.61 eV). A similar stabilization is maintained at 2.50% strain (−3.97 eV), indicating that moderate lattice expansion strengthens Li binding at the vacancy site. Despite these energetic changes, the electronic character of Li insertion remains essentially invariant. The Bader charge of Li stays close to +0.90 e across the investigated strain range, confirming persistent ionic electron donation to the defect manifold. The reduction of V shows only modest strain sensitivity, with Δq(V) increasing from −0.145 e at zero strain to −0.184 e at 1.25% strain and remaining comparable at 2.50% strain. Thus, moderate tensile strain slightly enhances dopant participation but does not qualitatively alter the vacancy-centered compensation mechanism.

**Table 2.** Effect of biaxial tensile strain applied in the (001) plane on Li insertion into the surface Ni vacancy of V-doped Ni-deficient NiO(001). Reported are the Li binding energy ($E_b$), the change in Bader charge of the V dopant upon lithiation ($\Delta q$(V)), and the Bader charge of the inserted Li atom ($q$(Li)) for different strain levels.

| Strain / % | $E_b$(Li) / eV | $\Delta q$(V) / e | $q$(Li) / e |
|---|---|---|---|
| 0 | −3.61 | −0.145 | +0.896 |
| 1.25 | −4.02 | −0.184 | +0.900 |
| 2.50 | −3.97 | −0.164 | +0.902 |

Biaxial tensile strain significantly affects the optical spectra of the non-lithiated V-doped Ni-deficient NiO(001) surface (**Figure 6**, top left), indicating a strain-induced disturbance in the defect-related electronic structure. As a result, the lithiated spectra also vary with the applied strain. For moderate tensile strain (1.25-2.50%), Li insertion continues to reduce visible absorption, indicating that the electrochromic mechanism remains active. However, the magnitude of the optical response decreases compared to that of the unstrained surface. This trend is clearly shown in the relative change of the absorption coefficient (**Figure 6**, top right), where the lithiation-induced decrease in absorption becomes smaller as tensile strain increases. Importantly, Bader analysis reveals that Li maintains a strongly ionic character across the evaluated strain range, and the change in V charge state remains small, indicating the fundamental charge-transfer mechanism remains intact. The reduced electrochromic contrast under strain does not stem from weakened electron donation but from strain-induced changes in the vacancy-related electronic structure in the non-lithiated surface. Tensile deformation alters the baseline distribution and strength of defect-derived optical transitions, limiting the extent to which lithiation can further reduce visible absorption. Thus, while moderate tensile strain improves Li binding energetics, it also decreases optical contrast by altering the defect states that drive electrochromic switching.

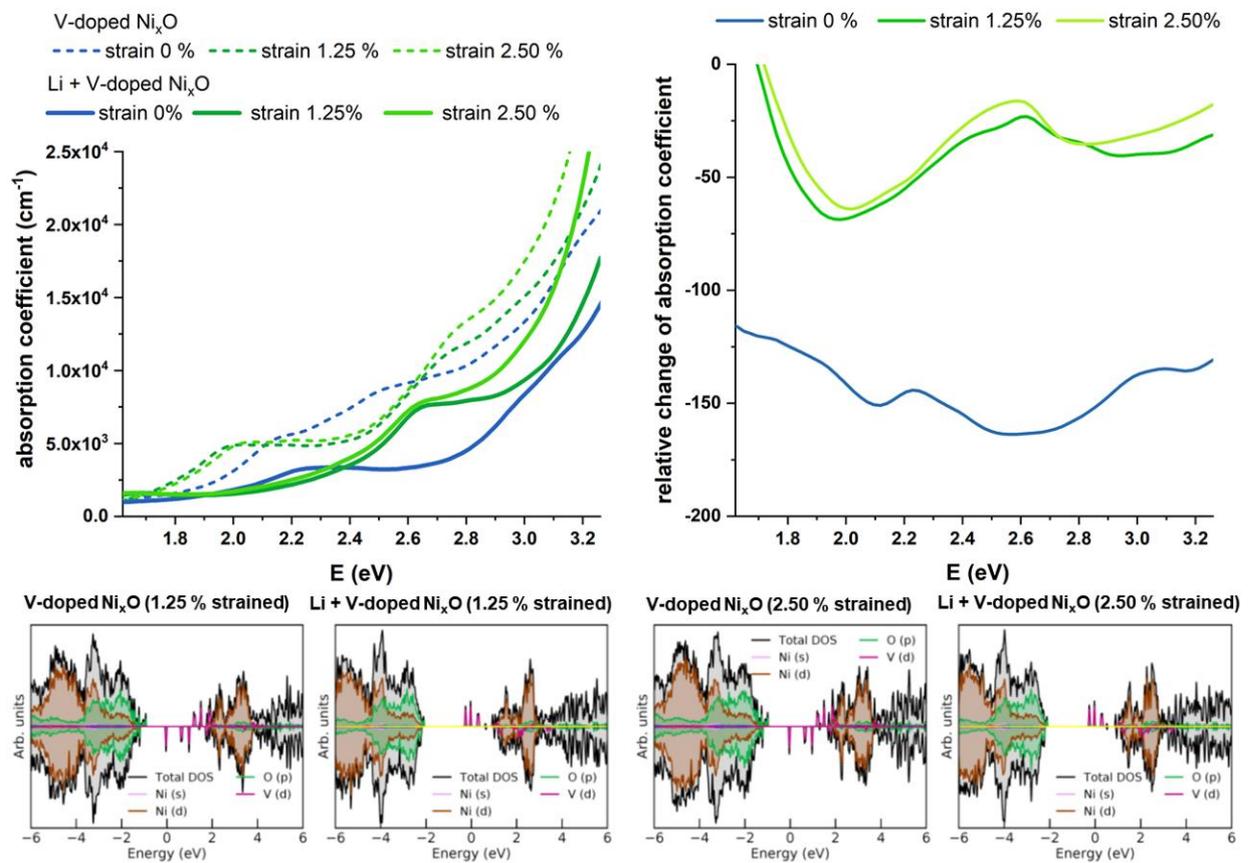

**Figure 6.** Top left: Calculated optical absorption spectra of V-doped Ni-deficient NiO(001) under biaxial tensile strain applied in the (001) plane. Dashed lines correspond to the non-lithiated surface at different strain levels (0%, 1.25%, and 2.50%), while solid lines represent the corresponding lithiated systems with Li inserted into the surface Ni vacancy. Top right: relative change of the adsorption coefficient upon the lithiation of the non-strained surface and the strained ones. Bottom row shows the total and projected density-of-states for the strained cases, before and after lithiation.

Biaxial strain represents a technologically relevant tuning parameter for electrochromic NiO thin films, which are commonly grown on lattice-mismatched substrates or subjected to mechanical constraints during device operation. Epitaxial mismatch, thermal expansion differences, and repeated electrochemical cycling can generate appreciable in-plane stress, particularly in ultrathin films. Because electrochromic switching in Ni-deficient NiO is mediated by defect states and local lattice distortions, even moderate strain can alter vacancy energetics, dopant–defect coupling, and band-edge alignment. In this context, strain engineering provides a potential route to modulate Li insertion thermodynamics and optical contrast without altering chemical composition. Our results show that moderate tensile strain (1-2.5%) enhances Li binding while preserving the vacancy-centered switching mechanism, indicating that electrochromic energetics can be tuned mechanically while maintaining robust bleaching behavior. This suggests that substrate selection and strain control can be used to optimize the switching voltage and stability of V-doped NiO-based electrochromic devices.

It is important to note that in realistic doped Ni-deficient NiO films, the defect landscape is unlikely to consist solely of idealized dopant-vacancy complexes. Besides dopant-associated Ni vacancies, isolated Ni vacancies and dopants without directly coordinated vacancies are expected to coexist. Each of these motifs represents a separate electronic center with its own charge-compensation pathway and optical signature. The present results show that isolated vacancy states promote conventional bleaching upon alkali insertion, while dopant-vacancy complexes can alter or even reverse the electrochromic response depending on dopant electronic activity. Therefore, the overall optical behavior of doped NiO films should be viewed as a superposition of contributions from multiple defect motifs, weighted by their relative populations. The balance between framework-dominated vacancy centers and dopant-coupled vacancy complexes offers an additional degree of freedom for tailoring electrochromic performance in NiO-based materials.

## 4. Conclusions

We have systematically investigated the electrochromic response of doped Ni-deficient NiO(001) surfaces by combining defect energetics, Bader charge analysis, density-of-states calculations, and optical absorption spectra. The results demonstrate that electrochromic switching in this system is governed by electron filling of vacancy-associated hole states, while the manner in which the injected electron is accommodated depends sensitively on dopant identity. Among the studied dopants, V preserves the framework-dominated compensation mechanism characteristic of undoped Ni-deficient NiO, resulting in a robust bleaching response upon alkali-ion insertion. In contrast, Sn actively participates in electronic compensation, trapping a significant fraction of the injected charge and reversing the optical response, whereas Cu behaves largely as a spectator, producing a non-monotonic spectral redistribution without substantial dopant reduction. These findings establish dopant-controlled partitioning of electron density between the oxide framework and dopant-centered states as the key factor determining electrochromic behavior. Alkali-ion identity (Li, Na, K) does not qualitatively alter the switching mechanism in the V-doped system. All three ions insert as strongly ionic donors and neutralize vacancy-derived hole states, leading to comparable bleaching behavior despite differences in binding strength and ionic size. This confirms that electron donation to the defect manifold, rather than specific cation chemistry, governs the optical response. Finally, biaxial tensile strain provides an additional, device-relevant tuning parameter. Moderate strain enhances Li binding energetics while preserving the vacancy-centered compensation mechanism, but progressively reduces the magnitude of the optical contrast by modifying the baseline defect electronic structure. Thus, electrochromic switching in doped Ni-deficient NiO emerges from a delicate interplay between defect energetics, dopant participation, and lattice deformation. This work establishes a unified mechanistic framework linking defect chemistry, dopant electronic activity, alkali insertion, and mechanical strain to the electrochromic response of NiO-based systems, offering clear design principles for strain- and dopant-engineered electrochromic materials.

**Acknowledgement**

A.S.D. and I.A.P. acknowledge the financial support provided by the Serbian Ministry of Science, Technological Development, and Innovations (contract no. 451-03-34/2026-03/200146). The computations and data handling were enabled by resources provided by the National Academic Infrastructure for Supercomputing in Sweden (NAISS) at the National Supercomputer center (NSC) at Linköping University, partially funded by the Swedish Research Council through grant agreements No. NAISS 2024/5-718 and NAISS 2025/5-713.

**References**

1   G. A. Niklasson and C. G. Granqvist, Electrochromics for smart windows: Thin films of tungsten oxide and nickel oxide, and devices based on these, *J. Mater. Chem.*, 2007, **17**, 127–156.

2   F. Zhao, T. Chen, Y. Zeng, J. Chen, J. Zheng, Y. Liu and G. Han, Nickel oxide electrochromic films: mechanisms, preparation methods, and modification strategies–a review, *J. Mater. Chem. C Mater.*, 2024, **12**, 7126–7145.

3   G. Boschloo and A. Hagfeldt, Spectroelectrochemistry of Nanostructured NiO, *Journal of Physical Chemistry B*, 2001, **105**, 3039–3044.

4   G. A. Niklasson and C. G. Granqvist, Electrochromics for smart windows: Thin films of tungsten oxide and nickel oxide, and devices based on these, *Journal of Materials Chemistry*, 2007, **17**, 127–156.

5   E. Avendaño, H. Rensmo, A. Azens, A. Sandell, G. de M. Azevedo, H. Siegbahn, G. A. Niklasson and C. G. Granqvist, Coloration Mechanism in Proton-Intercalated Electrochromic Hydrated $NiO_y$ and $Ni_{1-x}V_xO_y$ Thin Films, *J. Electrochem. Soc.*, 2009, **156**, P132.

6   R.-T. Wen, C. G. Granqvist and G. A. Niklasson, Anodic Electrochromism for Energy-Efficient Windows: Cation/Anion-Based Surface Processes and Effects of Crystal Facets in Nickel Oxide Thin Films, *Adv. Funct. Mater.*, 2015, **25**, 3359–3370.

7   R. T. Wen, C. G. Granqvist and G. A. Niklasson, Cyclic voltammetry on sputter-deposited films of electrochromic Ni oxide: Power-law decay of the charge density exchange, *Appl. Phys. Lett.*, 2014, **105**, 163502.

8   I. A. Pašti, A. S. Dobrota, D. B. Migas, B. Johansson and N. V. Skorodumova, Theoretical analysis of electrochromism of Ni-deficient nickel oxide – from bulk to surfaces, *Physical Chemistry Chemical Physics*, 2023, **25**, 7974–7985.

9   J. H. Zhang, G. F. Cai, D. Zhou, H. Tang, X. L. Wang, C. D. Gu and J. P. Tu, Co-doped NiO nanoflake array films with enhanced electrochromic properties, *J. Mater. Chem. C Mater.*, 2014, **2**, 7013–7021.

10  X. Zhan, F. Gao, Q. Zhuang, Y. Zhang and J. Dang, Two-Dimensional Porous Structure of V-Doped NiO with Enhanced Electrochromic Properties, *ACS Omega*, 2022, **7**, 8960–8967.

11  X. Yao, S. Ding, X. Shen, C. Guo, Y. Liu, W. Xia, G. Wu and Y. Zhang, Enhanced Electrochromic Properties of NiOx Films Through Magnesium Doping Strategy, *Nanomaterials 2025, Vol. 15,* 2025, **15**, 1217.

12  L. Zhang, C. Pan, H. Zeng, K. Han and Y. Gao, Amorphous and porous C/N-doped NiO for electrochromic smart windows applications, *Opt. Mater. (Amst).*, 2025, **167**, 117338.

13  I. Naskar, S. Roy, P. Ghosal and M. Deepa, Long-Lasting Panchromatic Electrochromic Device and Energy-Dense Supercapacitor Based on Zn-Doped NiO Microstars and a Redox-Active Gel, *ACS Appl. Energy Mater.*, 2023, **6**, 2385–2400.

14  Y. Zhang, X. Wang, G. Wei, R. Li, L. Jia, T. Li and M. Li, Synergistic enhancement of electrochromism and cyclic stability in Zn-doped NiO films via ethyl cellulose and In/Sn metal powder additives for large-area fabrication, *Electrochim. Acta*, 2026, **549**, 148009.

15  G. Kresse and J. Hafner, Ab initio molecular dynamics for liquid metals, *Phys. Rev. B*, 1993, **47**, 558–561.

16  G. Kresse and J. Furthmüller, Efficiency of ab-initio total energy calculations for metals and semiconductors using a plane-wave basis set, *Comput. Mater. Sci.*, 1996, **6**, 15–50.

17  G. Kresse and J. Furthmüller, Efficient iterative schemes for ab initio total-energy calculations using a plane-wave basis set, *Phys. Rev. B*, 1996, **54**, 11169–11186.

18  G. Kresse and D. Joubert, From ultrasoft pseudopotentials to the projector augmented-wave method, *Phys. Rev. B Condens. Matter Mater. Phys.*, 1999, **59**, 1758–1775.

19  J. Perdew, Burke and Ernzerhof, Generalized Gradient Approximation Made Simple., *Phys. Rev. Lett.*, 1996, **77**, 3865–3868.

20  P. E. Blöchl, Projector augmented-wave method, *Phys. Rev. B*, 1994, **50**, 17953–17979.

21  A. I. Liechtenstein, V. I. Anisimov and J. Zaanen, Density-functional theory and strong interactions: Orbital ordering in Mott-Hubbard insulators, *Phys. Rev. B*, 1995, **52**, R5467–R5470.

22  H. J. Monkhorst and J. D. Pack, Special points for Brillouin-zone integrations, *Phys. Rev. B*, 1976, **13**, 5188–5192.

23  P. E. Blöchl, O. Jepsen and O. K. Andersen, Improved tetrahedron method for Brillouin-zone integrations, *Phys. Rev. B*, 1994, **49**, 16223–16233.